# Single Step Phase Optimisation for Coherent Beam Combination using Deep Learning

Ben Mills[1]*, James A. Grant-Jacob[1], Matthew Praeger[1], Robert. W. Eason[1], Johan Nilsson[1], and Michalis N. Zervas[1]

Optoelectronics Research Centre, University of Southampton, SO171BJ, UK.

*b.mills@soton.ac.uk

## Abstract

Coherent beam combination of multiple fibres can be used to overcome limitations such as the power handling capability of single fibre configurations. In such a scheme, the focal intensity profile is critically dependent upon the relative phase of each fibre and so precise control over the phase of each fibre channel is essential. Determining the required phase compensations from the focal intensity profile alone (as measured via a camera) is extremely challenging with a large number of fibres as the phase information is obfuscated. Whilst iterative methods exist for phase retrieval, in practice, due to phase noise within a fibre laser amplification system, a single step process with computational time on the scale of milliseconds is needed. Here, we show how a neural network can be used to identify the phases of each fibre from the focal intensity profile, in a single step of ~ 10 milliseconds, for a simulated 3-ring hexagonal close packed arrangement, containing 19 separate fibres and subsequently how this enables bespoke beam shaping. In addition, we show that deep learning can be used to determine whether a desired intensity profile is physically possible within the simulation. This, coupled with the demonstrated resilience against simulated experimental noise, indicates a strong potential for the application of deep learning for coherent beam combination.

1. Introduction

High-power fibre lasers [1, 2] have seen increasing numbers of applications over the past few decades, as their maximum output power has risen by orders of magnitude during this time [3, 4]. Advancing beyond kilowatt average power levels, however, is challenging since high-power fibre lasers generally operate with a small mode field size and large propagation length, and hence, are subject to a range of nonlinear effects, such as Stimulated Raman Scattering [5], Stimulated Brillouin Scattering [6], and the optical Kerr effect [7]. Although the mode area can be increased, this reduces the threshold for transverse mode instability. An increasingly important technique for enhancing the total average power without increasing the average power of a single fibre is through the combination of multiple fibres emitting spectrally narrow co-aligned colinearly polarised beams, generally known as coherent beam combination [8-11]. However, an important consequence of combining multiple fibres is that changes in their relative phase will modify the resultant spatial intensity profile. Therefore, random differential phase fluctuations that are introduced externally or are intrinsic to the fibre laser amplification process, which might not be problematic when using a single fibre, are now a fundamentally limiting complication. There is therefore great interest in devising methods for identifying and correcting the phase of each fibre, in real-time, during operation. In practice, where the phase of each fibre can be accurately identified, a correcting signal could be sent to the phase actuator associated with each fibre, hence, resulting in a corrected focussed intensity profile.

Whilst there are methods for identification of the fibre phases through spatial interference or temporal beating [12, 13], a potentially more robust and lower cost solution is possible via analysis of the focal intensity profile. An added attraction of this is that it works directly with the targeted property, i.e., a desired beam profile. In practice, the focal intensity profile could be recorded on a camera, through using a beam splitter after the fibre array and focussing lens, on a separate beam path to that used for materials processing. The disadvantage of this relatively simple approach would be that the recorded focal intensity profile cannot directly provide information regarding the phase of each fibre, as only the modulus-squared of the electric field is recorded. Iterative phase retrieval methods exist, such as Gerchberg-Saxton and hybrid input-output (HIO) [14] algorithms. Other methods such as hill-climbing [15, 16] (e.g., stochastic parallel gradient descent, SPGD [17]) are often used, but require that the power is focused into a single spot. However, such methods are iterative, and in practice, a single-step, low-latency process would clearly be preferable.

Deep learning has recently seen a wide range of applications across laser optimisation [18, 19] and laser materials processing [20, 21], where neural networks have been shown to be as effective, or more effective, than traditional modelling approaches [22]. Critically, deep learning has the fundamentally important advantage of computation speed, as neural network implementations typically take just tens of milliseconds per calculation. Deep learning is particularly useful when the underlying processes are challenging, or impossible, to describe mathematically [23], and hence, neural networks are a natural choice for solving the challenge of reconstructing phases from an intensity profile, as shown in the general area of coherent diffractive imaging [24, 25].

Deep learning has also started to be applied to coherent beam combination, where recent results include the use of a neural network combined with an iterative gradient descent methods [26], processing of interference patterns [27], and reinforcement learning approaches [28-30], for the identification and control of phases. However, the fundamentally important milestone of accurate single-step determination of individual fibre phases in a multi-fibre coherent combination scheme has yet to be demonstrated. In this work, firstly we demonstrate that a neural network can identify and optimise all fibre phases in a single step. Secondly, we demonstrate that the neural network can be trained to select fibre phase corrections that target bespoke intensity profiles. Thirdly, we show an elegant application of two neural networks, which can be used to determine whether an arbitrary intensity profile is physically possible for phase combinations in a given fibre configuration.

2. Results

The primary objective of this work was to demonstrate a method for the identification of the phase profile in a fibre array, when only observing the intensity profile at, or near, the focus. Fundamentally, the challenge of phase identification arises due to the nature of measuring the intensity profile, where the phase information is obfuscated as only the intensity associated with the electric field is recorded. Therefore, whilst the transformation from known intensity and phase for each fibre, to the resultant focussed intensity profile is mathematically trivial, the inverse transformation from a measured intensity profile back to the intensity and phase of individual fibre phases generally requires a more complex approach, such as iterative algorithms (even when each fibre intensity is assumed to be constant). In this work, we demonstrate that by setting up the challenge of the identification of the phase of each fibre as a two-dimensional phase retrieval problem, we can identify an accurate solution, in a single step, using deep learning. Critically, our approach has additional constraints that eliminate two of the trivial ambiguities associated with two-dimensional phase retrieval, namely a global phase shift and a spatial shift. Firstly, a global phase shift, where an addition of a phase value across the whole image will provide an identical solution, is not possible in this case, as the phase of the central fibre is always set to an absolute phase value of zero. Second, a spatial shift, where the reconstructed field can be spatially shifted but with a compensatory linear phase gradient, is also not a possible solution in this case, as the position, amplitude and polarisation of all fibres are fixed.

Here, the deep learning approach uses a conditional generative adversarial network [31] that can transform an image associated with one domain into an image associated with another domain. We therefore use the neural network for transforming spatial intensity profiles into spatial phase profiles, hence acting as a single step phase retrieval calculator. Whilst deep learning eliminates the need for a physical understanding of a transformation, in practice the scientific effort is transferred to the production of suitable training data. The method for creating training data for this work is shown in figure 1, where randomly generated phases and a fixed intensity profile are applied to nineteen fibres formed into a hexagonal close-packed array, and the associated focused intensity profile is calculated via beam propagation mathematics [32] with the addition of a curved phase to simulate the effect of the focussing lens. As illustrated in the figure, a notable adaptation was to describe the phase using sine and cosine functions in two separate colour channels, to provide a cyclical colour change associated with the phase. A set of twelve randomly chosen training pairs are shown in the inset b) of the figure. A total of $3\times10^5$ training pairs and 1500 testing pairs were created.

The motivation for identification of the phase profile of the fibre array from the intensity profile is for demonstrating the concept of firstly phase correction, and secondly bespoke beam shaping. Here, we do not discuss the engineering challenges, such as beam alignment, associated with phase control of multiple individual optical fibres, and instead our focus is the concept of phase identification and correction. As shown in figure 2, the neural network is capable of the identification of the phase profile from the intensity profile (note the similarity between the current i.e. actual, and predicted phase profiles). When this predicted phase is subtracted from the current phase (which is both random and unknown to the neural network), the result is the creation of a flat phase profile (labelled subtracted phase). In some applications the subtracted phase may offer the desired output, however, this can optionally be augmented with a bespoke phase profile (i.e. the target phase can be added in order to produce the corrected phase output). In practice, one could envisage that this phase subtraction and phase addition could be managed in a single step via direct control of the phase actuator of each fibre. Whilst the figure shows the transformation into a 6-fold ring intensity profile, in practice, the end result could be any physically possible spatial intensity profile.

3. Discussion

Whilst here the neural network was provided with training data that corresponded to a theoretically perfect simulation, the resilience of the neural network to noise within the intensity profiles (i.e. camera images) is important to quantify. For this work, we introduced a unit of simulated experimental noise, which corresponded to a normal distribution with mean zero and standard deviation as the square root of the magnitude of the intensity profile, and which was chosen to be a realistic model of noise one might observe experimentally from a camera. As the simulated intensity profiles were converted to a grayscale 8-bit image file for use with the neural network (see figure 1), the possible intensity values for each image pixel varied between 0 and 255. Therefore, for one unit of experimental noise, a pixel value of 100 would generally vary between 90 and 110 (i.e. approximately 68% of values would fall in this range). The neural network was tested with 1, 10 and 100 units of simulated experimental noise, and a comparison made of the predictive accuracy. As can be seen in figure 3, the effect of noise at 1 and 10 units was minimal. The observation that the neural network was robust against a level of noise that could be typical in an experiment is perhaps surprising, given that the network was only trained on simulations without noise. Also shown in figure 3, are examples of predicted phase profiles for different numbers of training data pairs. There is a clear improvement in predictive capability as the number of training pairs is increased.

The effect of the amount of training data and degree of simulated noise on the neural network predictions are evaluated in more detail in figure 4, where predictions from 1500 test examples are presented. A key metric here is achieved using an arbitrarily chosen boundary, generally referred to as a bucket, which corresponds in this case to the approximate size of the central spot when a flat phase profile is chosen. The power in the bucket, i.e. the amount of intensity that is contained within the chosen region, can then be used to provide a quantitative comparison of the neural network predictive capability under different conditions. Here, the power in the bucket for any intensity profile is defined as the percentage of intensity contained within the bucket divided by the percentage of intensity contained within the bucket for a flat phase profile. Hence, in this work, the power in the bucket percentage metric is a measure of comparison with the "perfect" flat phase profile. The figure shows that for $10^3$ training pairs or fewer, the neural network predictive capability is no better than a random guess. Likewise, for more than $10^5$ training pairs, almost no further improvement is demonstrated. The cumulative distribution shows that increasing the amount of training data from $10^4$ to $10^5$ increases the percentage of test data that achieves 90% power in the bucket from 23% to 93%. This result implies that when applying deep learning to a phase retrieval problem, there may be a significant improvement in accuracy achieved through additional training data, but equally there may be a saturation point where returns diminish and additional training data no longer improves the accuracy. The cumulative distribution associated with simulated experimental noise shows clearly that the neural network predictive capability is only slightly affected by the inclusion of noise, especially if the magnitude of that noise is relatively small.

As illustrated earlier in figure 2, where a random intensity profile is modified into a ring pattern in a single step, a significant advantage of using an array of fibres is the prospect for real-time bespoke beam shaping. However, due to challenges associated with transforming an intensity profile into a phase profile, there is generally no direct way of knowing whether a desired intensity profile is physically possible. However, as shown in figure 5, through an innovative application of two neural networks, a single step process can be devised to test the validity of any desired intensity profile. To achieve this, a second neural network is trained that can perform the reverse operation, i.e. to transform a phase profile into an intensity profile. Whilst such a transformation is mathematically trivial, there is still a computational cost associated with this calculation, which can generally be reduced through the application of a neural network. To determine whether an intensity profile is physically possible (under the conditions simulated in this work), the intensity profile is passed through the first network, and the predicted phase profile is passed through the second network, which then predicts a second intensity profile. If the output of the second neural network is equal to the input of the first neural network, then the input intensity profile is possible. If the output intensity is different to the input intensity, then the input intensity is not possible. This effect can be understood by the relationship, within this simulation, where all possible phase profiles lead to an intensity profile, but not all intensity profiles lead to a phase profile (e.g. a square focussed intensity profile is not possible due to the fixed intensity distribution assumed at the fibre output). Using two neural networks in sequence is therefore a test of cyclic consistency between the two domains. Figure 5 shows the case for an intensity pattern (from the test data) corresponding to a 6-fold ring, where the original, a 30-degree rotation of the original, and a 60-degree rotation of the original, are tested for cyclicity. Due to the 6-fold rotational symmetry of the fibre array coordinates in this simulation, a 6-fold ring rotated by 30 degrees will not be physically possible. The neural network cyclicity test confirms this, by showing a significant difference between the input intensity and the output intensity. As expected, the 60-degree rotation is determined to be valid.

4. Conclusions

In conclusion, a method for predicting the phase of nineteen fibres arranged in a hexagonal close-packed array directly from the simulated focal intensity was shown, which has direct application in the optimisation of coherent beam combination. The approach used a conditional generative adversarial network to transform an image of the simulated focal intensity profile into the associated image of the simulated phase profile at the exit of the fibre array. It was shown that subtracting the predicted phase from the current phase would produce a good estimate of a flat phase, which could be used for phase correction or as a basis for adding bespoke phase profiles, and hence, enabling spatial intensity profile control. By training a second neural network to perform the inverse operation, namely the transformation of simulated phase into simulated intensity profile, the two networks could be linked, such that an intensity profile could be transformed into a phase profile and then back to an intensity profile. As, in this simulation, all possible phase profiles lead to an intensity profile, but not all intensity profiles lead to a phase profile, the two networks could be used to identify which intensity profiles were physically possible in this simulation.

5. Methods

**Beam propagation simulation:** The simulated data was created via the use of the angular spectrum method to propagate electric fields from the fibre exit plane to the focal plane. The simulated electric field was a 1000×1000 array, with pixel size of 10μm, a distance between the fibre plane and the focal plane of 25cm, a radius of each fibre of 500μm, and a laser wavelength of 1μm. The spatial distribution of the electric field amplitude for each fibre was a Gaussian with $1/e^2$ intensity radius as 0.8 of the fibre radius, with a maximum value of one, and with zero amplitude outside the fibre. The phase for each fibre was randomly chosen from a uniform distribution between -π and + π, and the phase for the central fibre was always set as zero. Each random set of phases therefore had an associated focal intensity profile, and hence, both could be used to create a single training (or testing) data pair for the neural network. It was found that a trigonometric representation of the phase was needed to create the phase image, as shown in figure 1, where the red channel of the image corresponded to the cosine of the phase, and the blue channel to the sine of the phase. This approach ensured that there was a cyclic change in the colour of the image, rather than a discrete jump from -π to + π. The intensity image was created by making all RGB channels equal to the simulated intensity value, and converted into an 8-bit number (i.e. 0 to 255), and hence acted as a similar quantisation of intensity values to an 8-bit monochrome experimental camera. For each intensity image, the values were normalised to a maximum of 255. Finally, to reduce training time, and hence allow training on a larger number of random phase combinations, the images were reduced to a 256×256 resolution. The fibre field was firstly cropped to 512×512, and then resized to 256×256. The focal intensity was cropped to 100×100 and then resized to 256×256. The simulated size scales are presented for the two domains in figure 1. Due to the computational challenge of creating large numbers of training data pairs needed for this work, a range of high specification personal computers were used (which typically created approximately 1000 training pairs per hour) as well as the IRIDIS High Performance Computing Facility at the University of Southampton.

**Neural network:** The network was a conditional generative adversarial network (cGAN), known in the literature as "pix2pix" [31]. This is a well-studied network, capable of learning complex transfer functions between images from two different domains, and hence ideal for learning the complex relationship between an intensity profile and its corresponding reconstructed phase profile. The generator had a U-Net structure, with a downscaling and upscaling path consisting of 8 blocks of 4×4 convolutional filters and strides of 2, each followed by a batch normalisation and a leaky ReLU [33]. The downscaling and upscaling paths were connected via concatenation. The discriminator consisted of downscaling using 4 blocks, also consisting of 4×4 convolutional filters and strides of 2, each followed by a batch normalisation and a leaky ReLU. A minibatch of size 1 was used, and a generator and discriminator learn rate of 0.0002. The neural network training ran for 1 epoch for all datasets. Training was completed using MATLAB, taking from 2 mins (300 epochs) to nearly 36 hours (300k epochs).

## 7. Acknowledgements

B.M. was supported by the Engineering and Physical Research Council Early Career Fellowships Scheme (EP/N03368X/1). M.N.Z. was supported by the Royal Academy of Engineering Research Chairs and Senior Research Fellowships Scheme. This work was supported by the Engineering and Physical Research Council under grant numbers EP/T026197/1 and EP/P027644/1. The authors acknowledge the use of the IRIDIS High Performance Computing Facility, and associated support services at the University of Southampton, in the completion of this work.

## 8. Author contributions

B.M. programmed the simulation for creating training data, analysed all neural network predictions and wrote the manuscript. J.A.G-J trained and optimised the neural networks. M.P. managed the training pair creation on the IRIDIS High Performance Computing Facility and provided analysis on the distribution of intensity profiles. R.W.E provided useful discussions on interpretation of the results. J.N. provided expertise in fibre lasers. M.N.Z. provided the motivation and direction of the investigation. All authors assisted with manuscript editing.

## 9. Additional Information

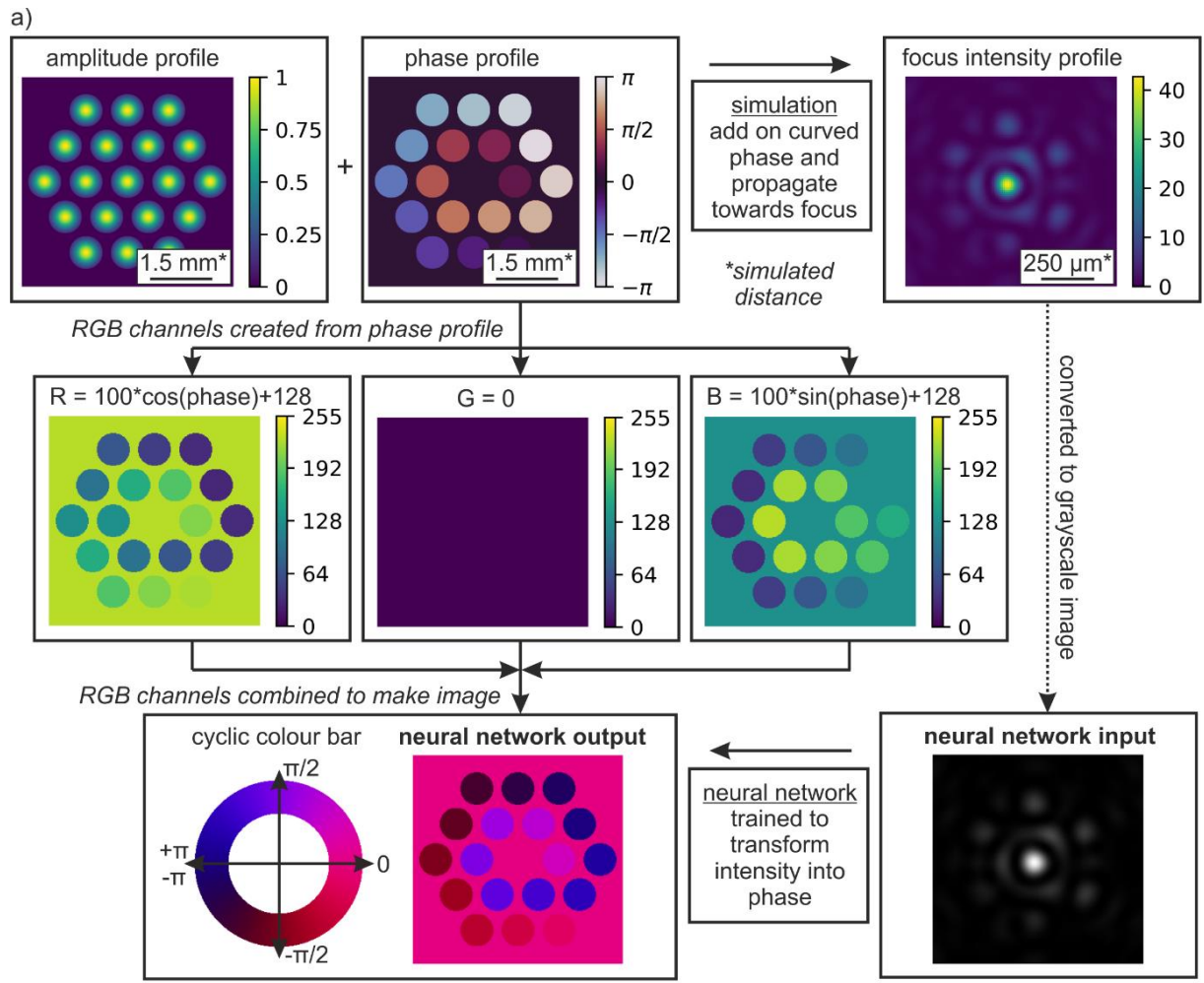

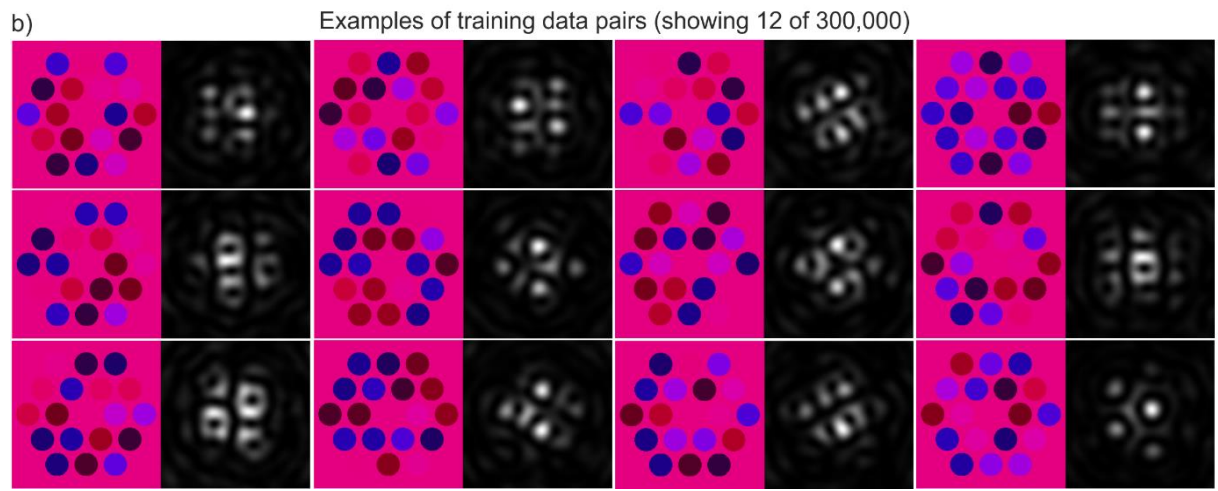

Fig. 1. Process for creation of training data suitable for training a neural network to transform a focal intensity profile into the associated phase profile that shows the phase of each fibre. Showing a) schematic of application of beam propagation simulation for creating neural network training data, which produces a 256×256 intensity image and associated 256×256 phase image. Showing b) twelve examples of training data pairs.

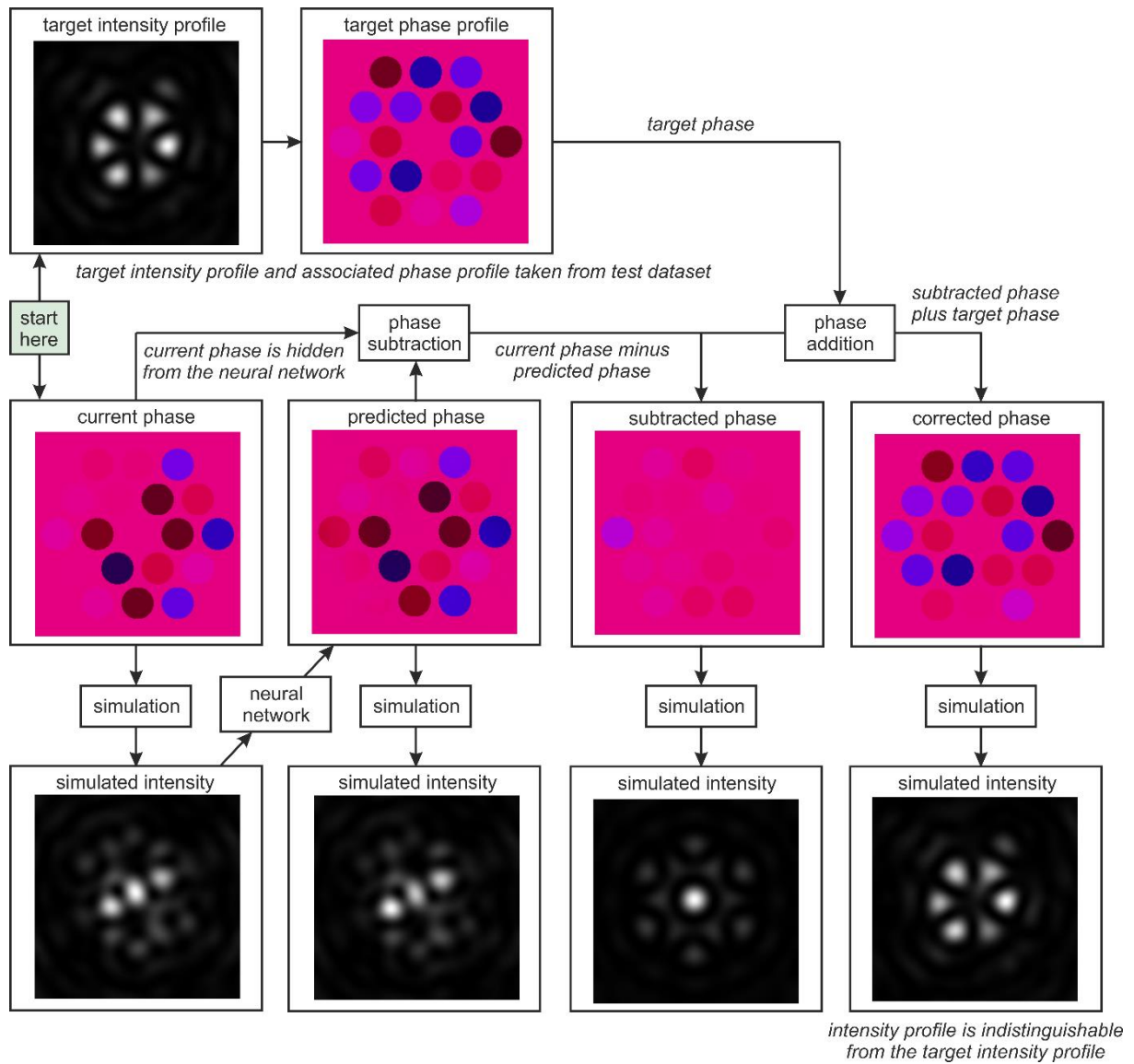

Fig 2. Application of a neural network for bespoke beam shaping for any phase that is unknown to the network. Starting from a current phase (which is unknown to the neural network), the associated simulated intensity is processed by the neural network and the phase is predicted. Subtracting the predicted phase profile from the (hidden) current phase profile produces a flat phase, with error depending on the prediction accuracy. At the same time, the phase profile for a desired intensity profile can be added to the corrected phase, to produce the desired intensity profile. This whole process is possible without knowledge of the current phase and could be completed in a single step.

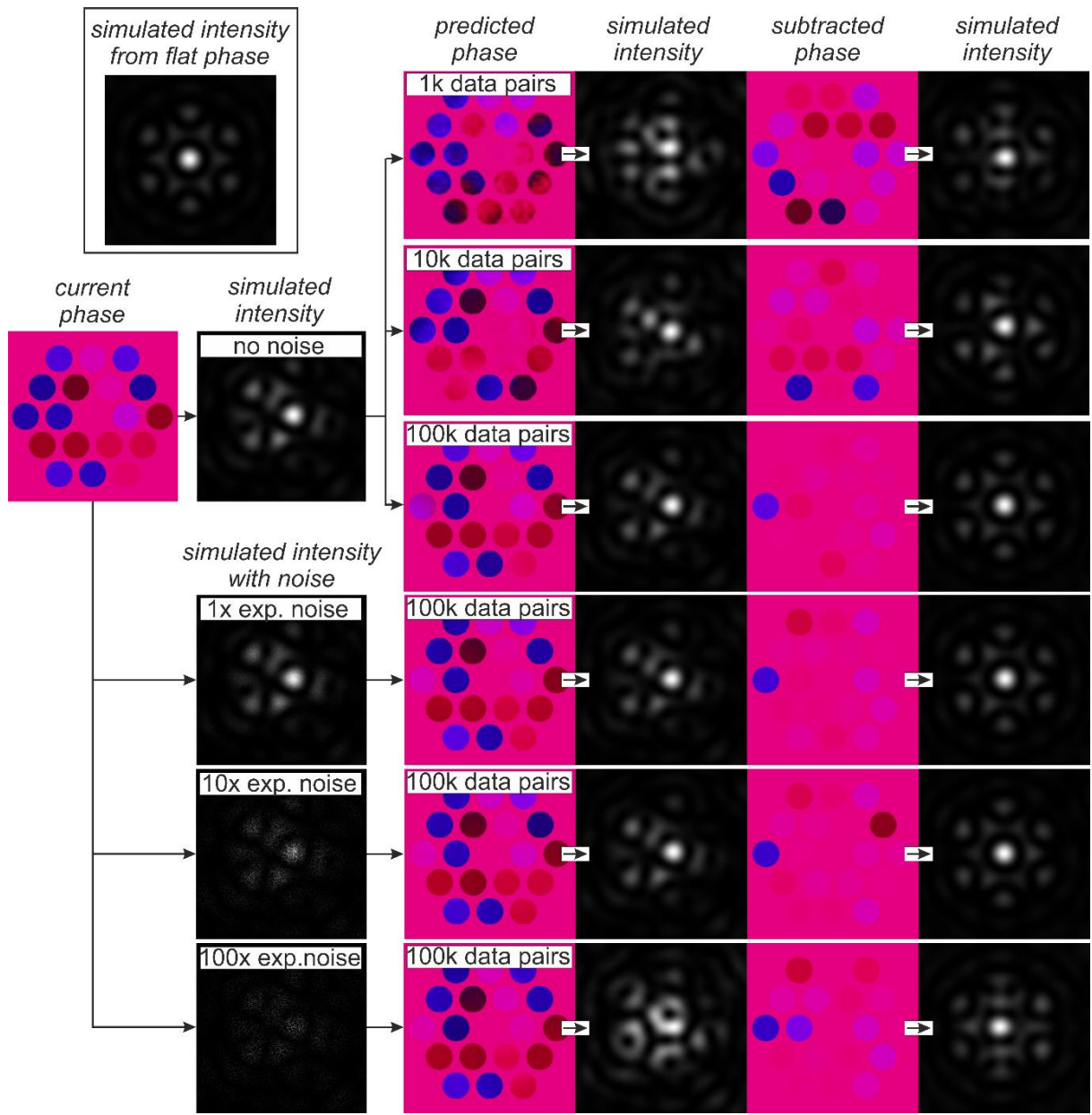

Fig 3. Neural network capability for different numbers of training pairs, and under different levels of simulated experimental noise. The results show clear improvements in the predictive capability as the number of training pairs is increased. The neural network is shown to be resilient under noise conditions that might be experimentally observed. Also shown, for comparison, is the intensity profile associated with a flat phase.

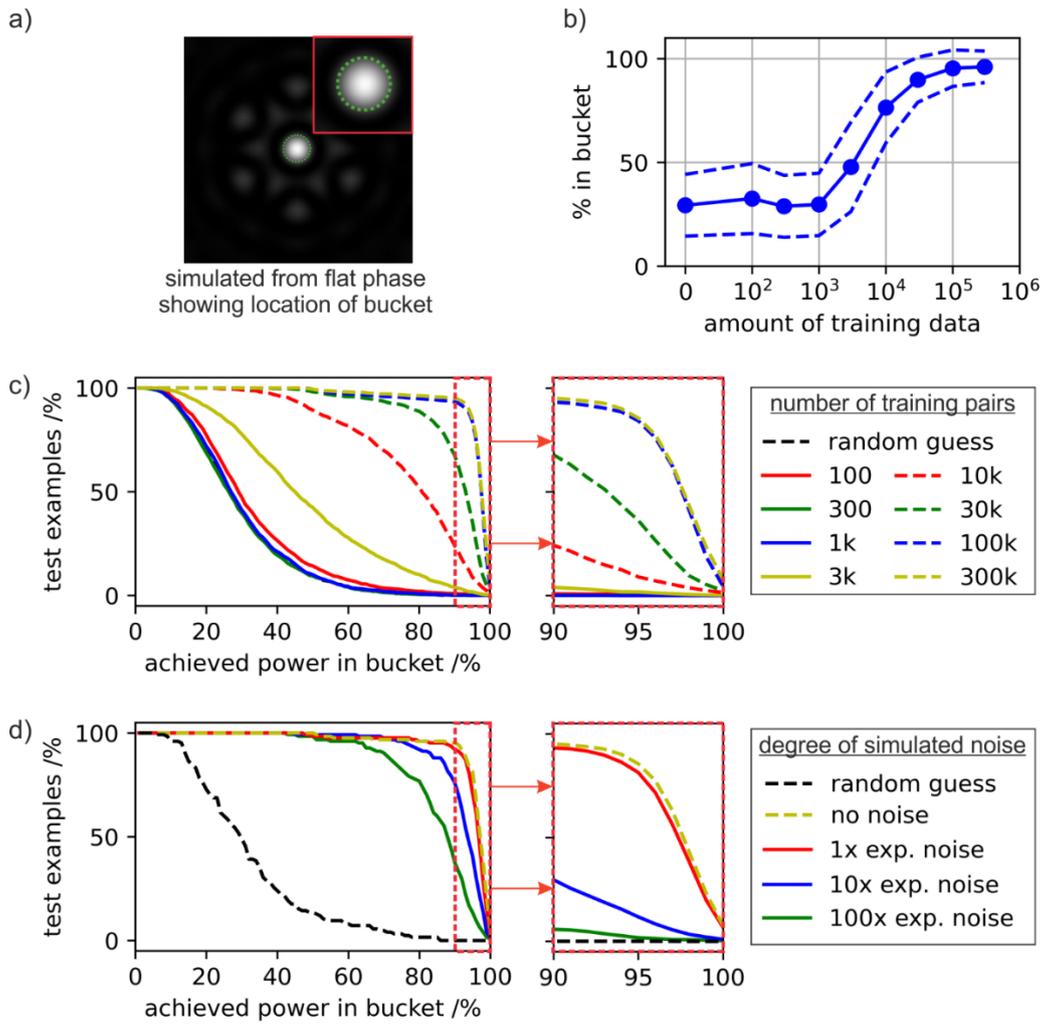

Fig. 4. Analysis of errors for 1500 randomly chosen test examples showing a) concept of power in the bucket, corresponding here to the percentage of intensity in the bucket for the corrected intensity profile divided by the percentage in the bucket for the flat phase case, b) the mean and standard deviation of power in the bucket for different numbers of training data pairs, c) distribution of test examples vs achieved power in the bucket for different amounts of training data and d) distribution of test examples vs achieved power in the bucket for different levels of simulated noise for the 300k case. Note that the line associated with the random guess is the same in both c) and d).

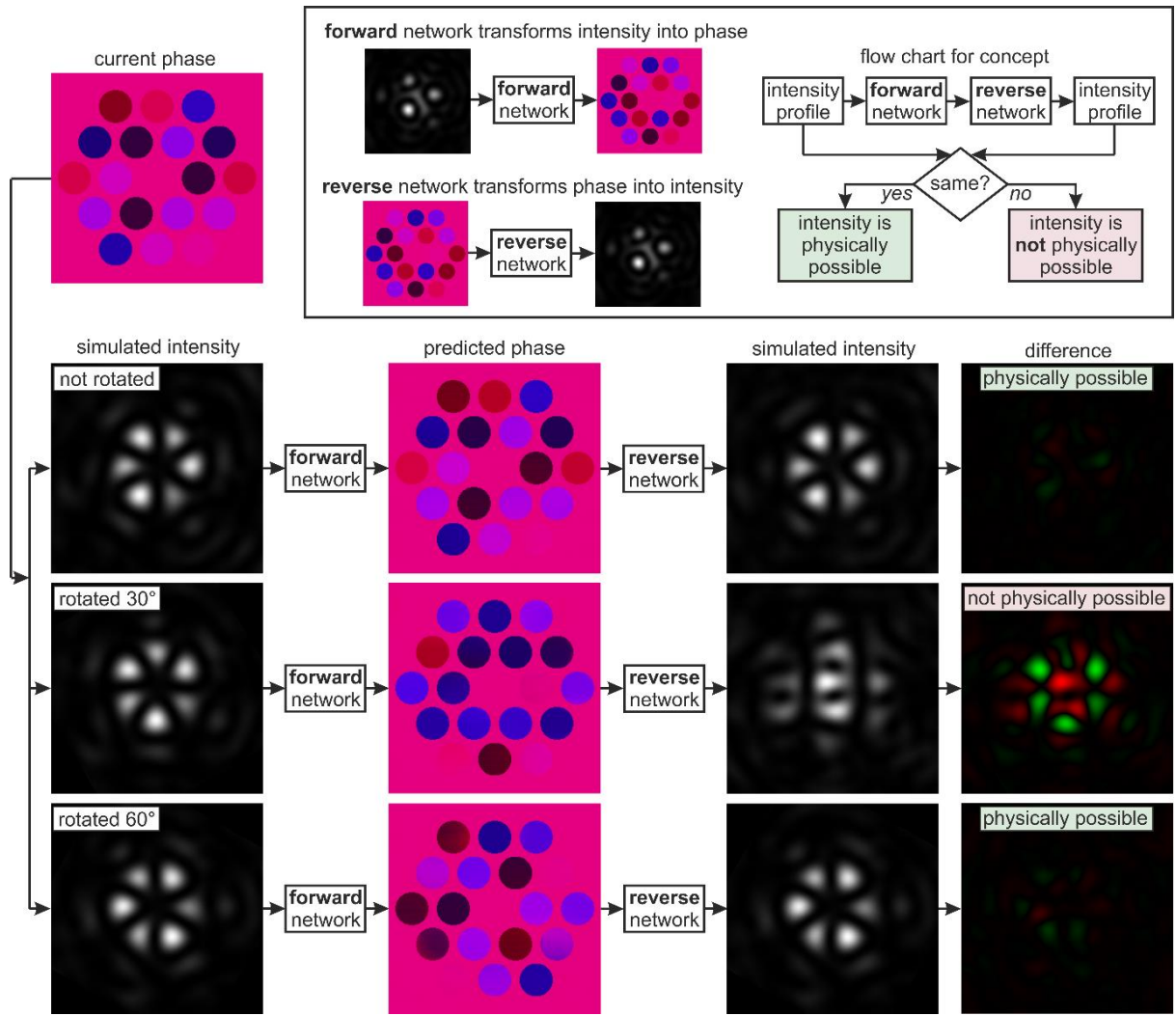

Fig. 5. Concept of using a forward and reverse neural network, in combination, for single step identification of whether an intensity profile is physically possible. In this case, the approach identifies whether a particular intensity profile is possible in this simulation. This method is possible as all simulated phases lead to simulated intensities, but not all simulated intensities lead to simulated phases. Due to the 6-fold rotational symmetry of the fibre array, a 30-degree rotation of the ring-shaped intensity profile is not a possible intensity profile, and hence there is no cyclicity between the domains (i.e. there is a difference between the input and output intensity profile). The difference column highlights where intensity is removed from the input intensity (green) or added to the input intensity (red), when comparing with the output intensity. The image pixel value for the red and green range from 0-255 according to the difference in pixel values for the input and output intensity profiles.